\documentclass[12pt,a4paper]{article}

\usepackage{amsmath}
\usepackage[varg]{txfonts}
\usepackage{cite}

\makeatletter
\renewcommand{\p@enumii}{}
\makeatother

\title{Extended Feynman Formula for the Harmonic Oscillator by the Discrete 
Time Method}
\author{Kunio FUNAHASHI\thanks{e-mail address : funahasi@isc.meiji.ac.jp}\\
  Department of Physics\\Meiji University\\Kawasaki 214-8571\\Japan}
\date{}

\begin{document}

\allowdisplaybreaks[4]

\maketitle

\begin{abstract}
  We calculate the Feynman formula for the harmonic oscillator beyond 
  and at caustics by the discrete formulation of path integral.
  The extension has been made by some authors, however, 
  it is not obtained by the method which we consider the most 
  reliable regularization of path integral.
  It is shown that this method leads to the result with,
  especially at caustics, more rigorous derivation than previous.
\end{abstract}

\section{Introduction}
\label{effhon-sec:introduction}

In optics it is known as the Guoy phase shift \cite{gouy} that 
when light waves pass through a focal point, the phase jumps $-\pi/2$
discontinuously.
This point is called the caustic point, in which the intensity of
light beam goes to infinity classically.
In quantum mechanics this is the point in which two or more
classical paths join.
So phenomena at caustics are observed in many systems which are in the
similar structure.
In path integral formula they make a change in the phase and this effect
is known as the Maslov correction \cite{maslov}.

The harmonic oscillator is one of the most important systems 
because it is solved exactly, it gives the first order approximations
of various systems and so on.
Path integral formula for the harmonic oscillator was obtained by 
Feynman himself \cite{feynmanhibbs}.
It is given by
\begin{equation}
  \label{effhon-eq:1}
  K(x_F,x_I;T)=\sqrt{\frac{m\omega}{2\pi i\hbar\sin\omega T}}
  \exp\biggl[i\frac{m\omega}{2\hbar\sin\omega T}\Bigl\{\bigl({x_I}^2+{x_F}^2
  \bigr)\cos\omega T-2x_Ix_F\Bigr\}\biggr]
\end{equation}
where $x_I$ ($x_F$) is the initial (final) position with the Hamiltonian
\begin{equation*}
  H=\frac1{2m}p^2+\frac12\omega^2x^2.
\end{equation*}
However it is known that this formula is valid only for a half-period.
In every half-period, infinite classical paths join at $x_I$ or $x_F$,
so the harmonic oscillator is the system in which caustics appear.
The extension for any time interval was made by Souriau \cite{souriau} first.
For $\omega T/\pi\ne[\omega T/\pi]$, it is given by
\begin{equation}
  \label{effhon-eq:2}
  K(x_F,x_I;T)=\sqrt{\frac{m\omega}{2\pi i\hbar|\sin\omega T|}}
  e^{-i\frac{[\omega T/\pi]}2\pi}
  \exp\biggl[i\frac{m\omega}{2\hbar\sin\omega T}\Bigl\{\bigl({x_I}^2+{x_F}^2
  \bigr)\cos\omega T-2x_Ix_F\Bigr\}\biggr]
\end{equation}
and for $\omega T/\pi=[\omega T/\pi]$, by
\begin{equation}
  \label{effho-eq:2}
  K(x_F,x_I;T)=e^{-i\frac{[\omega T/\pi]}2\pi}\,\delta(x_F-(-1)^{[\omega T/\pi]}x_I)
\end{equation}
where $[x]$ means the maximum integer not greater than $x$.
Later Horv\'athy \cite{horvathy} derived them by modifying Feynman's
original method and Liang and Morandi \cite{liangmorandi} by using 
the eigenfunction expansion for the propagator.
However there seems to be no derivation by the discrete time formulation
of path integral (the discrete time method).
Therefore we derive the extended Feynman formula by this method.

Path integral is a powerful quantization method and has plain idea,
however, in practical calculation some difficulties arise, for example,
integral measure is not necessarily given.
The discrete time method is one of the regularizations of path integral 
and in some cases it is indispensable \cite{su2,cpn,grassmann}.
The outline of this formulation is as follows.
Path integral formula is defined by
\begin{equation*}
  K(x_F,t_F;x_I,t_I)=\langle x_F|e^{-\frac i\hbar\hat H(t_F-t_I)}|x_I\rangle
\end{equation*}
or, if the Hamiltonian has time translation invariance, it is given by
\begin{equation}
  \label{effho-eq:1}
  K(x_F,x_I;T)=\langle x_F|e^{-\frac i\hbar\hat HT}|x_I\rangle
\end{equation}
where $T$ is the time interval.
Then we write (\ref{effho-eq:1}) as
\begin{align*}
  K(x_F,x_I;T)&=\lim_{N\to\infty}\langle x_F|(1-\frac i\hbar\Delta t\hat H)^N|x_I\rangle\\
  &=\lim_{N\to\infty}\int\prod_{i=1}^{N-1}dx_i\prod_{j=1}^N
  \langle x_j|(1-\frac i\hbar\Delta t\hat H)|x_{j-1}\rangle
\end{align*}
where we have successively inserted the completeness relation
\begin{equation*}
  \int dx_i|x_i\rangle\langle x_i|=1
\end{equation*}
and put $\Delta t=T/N$, $x_j=x(j\Delta t)$, $x_N=x_F$, $x_0=x_I$.
In the harmonic oscillator, by making use of 
\begin{equation*}
  \langle x_j|p_j\rangle=\int\frac{dp_j}{\sqrt{2\pi\hbar}}e^{\frac i\hbar p_jx_j},
\end{equation*}
the matrix element is given by
\begin{align}
  \langle x_j|(1-\frac i\hbar\Delta t\hat H)|x_{j-1}\rangle
  &=\int\frac{dp_j}{2\pi\hbar}\langle x_j|p_j\rangle\langle p_j|x_{j-1}\rangle
  \biggl[1-\frac i\hbar\Delta t\Bigl(\frac{{p_j}^2}{2m}+\frac m2\omega^2
  {\overline{x}_j}^2\Bigr)\biggr]\notag\\
  &=\int\frac{dp_j}{2\pi\hbar}e^{\frac i\hbar p_j\Delta x_j}
  \exp\biggl[-\frac i\hbar\Delta t\Bigl(\frac{{p_j}^2}{2m}+\frac m2\omega^2
  {\overline{x}_j}^2\Bigr)\biggr]+\mathrm{O}(\Delta t^2).
  \label{effhon-eq:5}
\end{align}
Then  carrying out the $p_j$-integrals, we obtain
\begin{gather}
  \label{effhon-eq:7}
  K(x_F,x_I;T)=\lim_{N\to\infty}\Bigl(\frac m{2\pi i\hbar\Delta t}\Bigr)^{N/2}
    \int\prod_{i=1}^{N-1}dx_i
    \exp\Bigl[\frac i\hbar S\Bigr],\\
    S=\Delta t\sum_{j=1}^N\biggl\{\frac m2\Bigl(\frac{\Delta x_j}{\Delta t}
    \Bigr)^2-\frac12m\omega^2{\overline{x}_j}^2\biggr\}
    \label{effho-eq:4}
\end{gather}
where we have put $\Delta x_j=x_j-x_{j-1}$, $\overline{x}_j=(x_j+x_{j-1})/2$
and omitted $\mathrm{O}(\Delta t^2)$ terms to disappear in $N\to\infty$.
This is the path integral formula for the harmonic oscillator by the
discrete time method.
Naive integration of (\ref{effhon-eq:7}) leads to the original Feynman
formula (\ref{effhon-eq:1}).
In \S\ref{effhon-sec:feynm-form-discr} we calculate (\ref{effhon-eq:7})
carefully to derive the extended Feynman formula (\ref{effhon-eq:2})
and (\ref{effho-eq:2}).

\section{The Extended Feynman formula}
\label{effhon-sec:feynm-form-discr}

At first we write (\ref{effho-eq:4}) by matrix notation as
\begin{gather}
  S=\frac m{2\Delta t}\bigl({}^t\!\boldsymbol xA\boldsymbol x+2{}^t\!
  \boldsymbol b\boldsymbol x+c\bigr)\label{effho-eq:5}\\
  A=\begin{pmatrix}
    2\alpha&-\beta&0&0&\cdots&0\\
    -\beta&2\alpha&-\beta&0&\cdots&0\\
    0&-\beta&2\alpha&&&\vdots\\
    0&0&&\ddots&&0\\
    \vdots&\vdots&&&2\alpha&-\beta\\
    0&0&\cdots&0&-\beta&2\alpha\end{pmatrix},\quad
  \boldsymbol b=-\beta{}^t\!(x_I,0,\ldots,0,x_F),\quad
  c=\alpha({x_I}^2+{x_F}^2)\notag\\
  \alpha=1-\frac{\omega^2\Delta t^2}4,\quad
  \beta=1+\frac{\omega^2\Delta t^2}4\notag
\end{gather}
where $t$ means transposition.
The $x_i$-integrals in (\ref{effhon-eq:7}) are the Fresnel integrals and 
they are evaluated by
\begin{equation}
  \label{effho-eq:7}
  \int_{-\infty}^\infty e^{i\alpha x^2}\,dx=e^{i\frac\pi4}\sqrt{\frac\pi\alpha},\quad
  \int_{-\infty}^\infty e^{-i\alpha x^2}\,dx=e^{-i\frac\pi4}\sqrt{\frac\pi\alpha}
  \qquad(\alpha>0).
\end{equation}
So we need to know numbers of positive eigenvalues and negative one.
By easy calculation we find the eigenvalues of $A$ are
\begin{equation}
  \label{effhon-eq:8}
  \lambda_k=2\bigl(\alpha-\beta\cos\frac{k\pi}N\bigr)\qquad(k=1,2,\ldots,N-1)
\end{equation}
or 
\begin{equation}
  \label{effhon-eq:10}
  \lambda_k=4\cos^2\frac{k\pi}{2N}\Bigl(-\frac{\omega^2\Delta t^2}4
  +\tan^2\frac{k\pi}{2N}\Bigr)
\end{equation}
without $\alpha$ and $\beta$ and the corresponding normalized eigenvectors are
\begin{equation}
  \label{effhon-eq:9}
  (\boldsymbol v_k)_l=\sqrt{\frac2N}\sin\frac{lk}N\pi\qquad(k,l=1,2,\ldots,N-1)
\end{equation}
where $(\boldsymbol v_k)_l$ means the $l$ th element of $\boldsymbol v_k$.

Since cosine decreases monotonously on $0\le x\le\pi$, we find $\lambda_k$'s
are ordered according to $k$.
So if we obtain the zero point $x_0(N)$ of
\begin{equation*}
  f(x)=2\bigl(\alpha-\beta\cos\frac\pi Nx\bigr)\in\mathrm{C}^0[0,N],
\end{equation*}
then $k$ less than $[x_0(N)]$ corresponds to negative eigenvalues.
Since $f(x)$ increases monotonously and $f(0)<0$ and $f(N)>0$, 
$f(x)=0$ has a unique solution
\begin{equation}
  \label{effho-eq:3}
  x_0(N)=\frac N\pi\cos^{-1}\frac\alpha\beta=\frac{2N}\pi\tan^{-1}\frac{\omega T}{2N}
\end{equation}
where we have used $\cos^{-1}x=2\tan^{-1}\sqrt{\frac{1-x}{1+x}}$.
Further by the Maclaurin expansion of arctangent, 
(\ref{effho-eq:3}) is evaluated by
\begin{equation*}
  x_0(N)=\frac{\omega T}\pi\biggl\{1-\frac13\Bigl(\frac{\omega T}{2N}\Bigr)^2
  +\cdots\biggr\}
\end{equation*}
so we find
\begin{equation}
  \label{effho-eq:6}
  x_0(N)\uparrow\frac{\omega T}\pi.
\end{equation}
Now we can count number of eigenvalues of each sign.
Two cases arise whether $\omega T/\pi$ is integer or not.
We put $M=[\omega T/\pi]$ for simplicity.
\begin{enumerate}
  \renewcommand{\theenumi}{\roman{enumi}}
  \renewcommand{\labelenumi}{(\theenumi)}
\item\label{effhon-item:1} $\omega T/\pi\ne M$
  
  Because $M<\frac{\omega T}\pi<M+1$, for sufficiently large $N$
  \begin{equation*}
    M<x_0(N)<M+1.
  \end{equation*}
  So $M$ of $N-1$ eigenvalues are negative and $N-M-1$ are positive.
\item\label{effhon-item:2} $\omega T/\pi=M$

  Because $x_0(N)<M$ and $x_0(N)\uparrow M$, for large but finite $N$,
  $M-1$ of $N-1$ eigenvalues are negative and $N-M$ are positive
  and in $N\to\infty$, $M-1$ remain negative and the $M$th goes to $0$ 
  and $N-M-1$ are positive.
\end{enumerate}
In both cases, for sufficiently large $N$, $\lambda_k\ne0$ and so $\det A\ne0$.
To deal with both cases together, we put
\begin{equation}
  \label{effhon-eq:12}
  L=
  \begin{cases}
    M&(\omega T\ne M\pi)\\
    M-1&(\omega T=M\pi)
  \end{cases}.
\end{equation}

We proceed to calculate (\ref{effho-eq:5}).
We put $\boldsymbol x_c$ as the solution of
$A\boldsymbol x+\boldsymbol b=\boldsymbol0$ :
$A\boldsymbol x_c+\boldsymbol b=\boldsymbol0$.
Then making use of  the translation 
$\boldsymbol x=\boldsymbol y+\boldsymbol x_c$, we obtain
\begin{equation*}
  S=\frac m{2\Delta t}\bigl({}^t\!\boldsymbol yA\boldsymbol y-{}^t\!\boldsymbol b
  A^{-1}\boldsymbol b+c\bigr).
\end{equation*}
Further making use of the orthogonal transformation 
$\boldsymbol y=P\boldsymbol z$ with 
$P=(\boldsymbol v_1,\boldsymbol v_2,\ldots,\boldsymbol v_{N-1})$
, we obtain
\begin{gather}
  \label{effhon-eq:11}
  K(x_F,x_I;T)=\lim_{N\to\infty}Q\exp\Bigl[\frac i\hbar S_c\Bigr]\\
  Q=\biggl(\frac m{2\pi i\hbar\Delta t}\biggr)^{N/2}
  \biggl\{\prod_{k=1}^{N-1}\int dx_k\,e^{\frac i\hbar\frac m{2\Delta t}\lambda_k{x_k}^2}\biggr\}
  ,\quad
  S_c=\frac m{2\Delta t}\bigl(-{}^t\!\boldsymbol bA^{-1}\boldsymbol b+c\bigr)\notag
\end{gather}
where we have rewritten $z$ to $x$.
First we calculate $Q$.
Taking account of the difference of the Frensel integrals (\ref{effho-eq:7})
by signs of eigenvalues, $Q$ is evaluated by
\begin{equation}
  \label{effho-eq:10}
  Q=\biggl(\frac m{2\pi i\hbar\Delta t}\biggr)^{N/2}
  \Biggl(\prod_{k=1}^L\int dx_k\,e^{-\frac i\hbar\frac m{2\Delta t}|\lambda_k|{x_k}^2}\Biggr)
  \Biggl(\prod_{k=L+1}^{N-1}\int dx_k\,e^{\frac i\hbar\frac m{2\Delta t}|\lambda_k|{x_k}^2}\Biggr)
  =\sqrt{\frac m{2\pi i\hbar\Delta t}}
  \frac{e^{-i\frac L2\pi}}{\sqrt{\prod_{k=1}^{N-1}|\lambda_k|}}.
\end{equation}
Making use of the well-known formulas
\begin{gather*}
  \prod_{r=1}^{n-1}\cos\frac{r\pi}{2n}=\frac{\sqrt{n}}{2^{n-1}},\quad
  \prod_{r=1}^{n-1}\Bigl(x^2+\tan^2\frac{rx}{2n}\Bigr)
  =\frac1{4nx}\Bigl[(x+1)^{2n}-(x-1)^{2n}\Bigr]
\end{gather*}
to (\ref{effhon-eq:10}), we immediately find
\begin{equation}
  \label{effho-eq:11}
  \prod_{k=1}^{N-1}|\lambda_k|=\frac N{\omega T}
  \biggl|\frac{\sigma^2(N)-{\overline{\sigma}}^2(N)}{2i}\biggr|
\end{equation}
where we have put $\sigma(N)=(1+i\frac{\omega T}{2N})^N$.
Putting (\ref{effho-eq:11}) into (\ref{effho-eq:10}), we obtain
\begin{equation}
  \label{effhon-eq:13}
  Q=\sqrt{\frac{m\omega}{2\pi i\hbar}}
  \frac{e^{-i\frac L2\pi}}{\sqrt{\Bigl|\frac{\sigma^2(N)-{\overline{\sigma}}^2(N)}
      {2i}\Bigr|}}.
\end{equation}

Next we calculate $S_c$.
We write the determinant of the ``$N-1$'' dimensional matrix $A$ as $D_{N-1}$.
Then $D_n$'s satisfy
\begin{equation*}
  D_{n+1}=2\alpha D_n-\beta^2 D_{n-1},\quad D_0=1,\quad D_1=2\alpha
\end{equation*}
so we find
\begin{equation*}
  D_{N-1}=\frac N{\omega T}\frac{\sigma^2(N)-{\overline{\sigma}}^2(N)}{2i}
\end{equation*}
and
\begin{equation*}
  A^{-1}=\frac1{D_{N-1}}
  \begin{pmatrix}
    D_{N-2}&\beta D_{N-3}&\cdots&\beta^{N-3}D_1&\beta^{N-2}\\
    *&*&\cdots&*&*\\
    \vdots&\vdots&\ddots&\vdots&\vdots\\
    *&*&\cdots&*&*\\
    \beta^{N-2}&\beta^{N-3}D_1&\cdots&\beta D_{N-3}&D_{N-2}
  \end{pmatrix}
\end{equation*}
where we have written ingredients which are unnecessary for later calculations 
as $*$.
Putting these into (\ref{effhon-eq:11}), we obtain
\begin{align}
  S_c&=\frac m{2\Delta t}\biggl\{\Bigl(\alpha-\beta^2\frac{D_{N-2}}{D_{N-1}}\Bigr)
  \bigl({x_I}^2+{x_F}^2\bigr)-\frac{\beta^N}{D_{N-1}}2x_Ix_F\biggr\}\notag\\
  &=i\frac{m\omega}{2\hbar}\frac{\bigl(\sigma^2(N)+{\overline{\sigma}}^2(N)
    \bigr)\bigl({x_I}^2+{x_F}^2\bigr)-\sigma(N)\overline{\sigma}(N)4x_Ix_F}
  {\sigma^2(N)-{\overline{\sigma}}^2(N)}.
  \label{effhon-eq:14}
\end{align}
By (\ref{effhon-eq:13}) and (\ref{effhon-eq:14}), the Feynman formula 
(\ref{effhon-eq:11}) is expressed as
\begin{align}  
  K(x_F,x_I;T)&=\lim_{N\to\infty}\sqrt{\frac{m\omega}{2\pi i\hbar}}
  \frac{e^{-i\frac L2\pi}}{\sqrt{\Bigl|\frac{\sigma^2(N)-{\overline{\sigma}}^2(N)}
      {2i}\Bigr|}}
  \times\exp\biggl[-\frac{m\omega}{2\hbar}
  \frac1{\sigma^2(N)-{\overline{\sigma}}^2(N)}\notag\\
  &\qquad\times\Bigl\{\bigl(\sigma^2(N)+{\overline{\sigma}}^2(N)\bigr)
  \bigl({x_I}^2+{x_F}^2\bigr)
  -\sigma(N)\overline{\sigma}(N)4x_Ix_F\Bigr\}\biggr].
  \label{effhon-eq:15}
\end{align}
Because $\sigma(N)\to e^{i\omega T/2}\ (N\to\infty)$, if 
$\omega T/2\in\mathbb{N}\cup\{0\}$
then $\sigma^2(N)-{\overline{\sigma}}^2(N)\to0\ (N\to\infty)$,
so we need to classify whether $\omega T/2\in\mathbb{N}\cup\{0\}$ or not.

\begin{enumerate}
  \renewcommand{\theenumi}{\roman{enumi}}
  \renewcommand{\labelenumi}{(\theenumi)}
\item $\omega T/\pi\not\in\mathbb{N}\cup\{0\}$
  
  In this case no divergence appears.
  Each factor converges:
  \begin{equation*}
    \sigma^2(N)-{\overline{\sigma}}^2(N)\to2i\sin\omega T,\
    \sigma^2(N)+{\overline{\sigma}}^2(N)\to2\cos\omega T,\
    \sigma(N)\overline{\sigma}(N)\to1\qquad(N\to\infty)
  \end{equation*}
  and $L=M$, so (\ref{effhon-eq:15}) is
  \begin{equation}
    \label{effhon-eq:16}
    K(x_F,x_I;T)=\sqrt{\frac{m\omega}{2\pi i\hbar|\sin\omega T|}}
    e^{-i\frac M2\pi}
    \exp\biggl[i\frac{m\omega}{2\hbar\sin\omega T}
    \Bigl\{\bigl({x_I}^2+{x_F}^2\bigr)\cos\omega T-2x_Ix_F\Bigr\}\biggr].
  \end{equation}
  This is in accordance with (\ref{effhon-eq:2}).
\item $\omega T/\pi\in\mathbb{N}\cup\{0\}$
  
  In this case $L=M-1$ and now divergence appears.
  To handle divergence we slightly rewrite (\ref{effhon-eq:15}) as
  \begin{align}  
    K(x_F,x_I;T)&=\lim_{N\to\infty}\sqrt{\frac{m\omega}{\pi i\hbar}}
    \frac{e^{-i\frac{M-1}2\pi}}{\sqrt{|\sigma(N)|^2}\sqrt{|1-z^2(N)|}}\notag\\
    &\qquad\times\exp\biggl[\frac{m\omega}{2\hbar}\bigl({x_I}^2+{x_F}^2\bigr)
    -\frac{\frac{m\omega}\hbar}{1-z^2(N)}\bigl({x_I}^2+{x_F}^2-2z(N)x_Ix_F
    \bigr)\biggr]
    \label{effhon-eq:17}
  \end{align}
  where we have put $z(N)=\overline{\sigma}(N)/\sigma(N)$ for simplicity.
  Further we put
  \begin{equation}
    \label{effhon-eq:18}
    u=\sqrt{\frac{m\omega}{2\hbar}}(x_I+x_F),\quad
    v=\sqrt{\frac{m\omega}{2\hbar}}(x_I-x_F),
  \end{equation}
  then we can rewrite (\ref{effhon-eq:18}) to
  \begin{equation}
    \label{effhon-eq:19}
    K(x_F,x_I;T)=\lim_{N\to\infty}\sqrt{\frac{m\omega}{\pi i\hbar}}
    \frac{e^{-i\frac{M-1}2\pi}}{\sqrt{|\sigma(N)|^2}}e^{\frac{u^2+v^2}2}
    \frac{e^{-\frac{u^2}{1+z(N)}}}{\sqrt{|1+z(N)|}}
    \frac{e^{-\frac{v^2}{1-z(N)}}}{\sqrt{|1-z(N)|}}.
  \end{equation}  
  For sufficiently large $N$
  \begin{equation*}
    \arg\sigma(N)=N\tan^{-1}\frac{M\pi}{2N}
    =\frac{M\pi}2-\frac13\Bigl(\frac{M\pi}2\Bigr)^3\frac1{N^2}+\cdots
  \end{equation*}
  so we can put
  \begin{equation*}
    \arg\sigma(N)=\frac{M\pi}2-\varepsilon(N),\quad
    \varepsilon(N)>0,\quad\varepsilon(N)\to0\ (N\to\infty)
  \end{equation*}
  and then we immediately obtain
  \begin{equation}
    \label{effho-eq:8}
    z(N)=e^{-2i\arg\sigma(N)}=(-1)^Me^{2i\varepsilon(N)}.
  \end{equation}
  By (\ref{effho-eq:8}) we find there are two cases of divergence in 
  (\ref{effhon-eq:19}).
  \begin{enumerate}
    \renewcommand{\theenumii}{\alph{enumii}}
    \renewcommand{\labelenumii}{(\theenumii)}
  \item\label{effhon-item:3} $M\in2\mathbb{N}\cup\{0\}$

    In this case
    \begin{equation*}
      1-z(N)=2\sin\varepsilon(N)\,e^{-i(\frac\pi2-\varepsilon(N))}
    \end{equation*}
    and
    \begin{equation*}
      z(N)\to1,\quad\arg(1-z(N))=-\frac\pi2\qquad(N\to\infty),
    \end{equation*}
    so the denominator and the numerator of the last factor in 
    (\ref{effhon-eq:19}) diverge.

    To handle this divergence we consider not $K(x_F,x_I;T)$ itself but
    \begin{equation}
      \label{effhon-eq:20}
      \int_{-\infty}^\infty K(x_F,x_I;T)f(x_I)\,dx_I
    \end{equation}
    where $f(x_I)$ is a well-behaved function because path integral formula
    has essentially meaning within integral.
    The explicit expression of (\ref{effhon-eq:20}) is
    \begin{align}
      (\ref{effhon-eq:20})
      &=\int_{-\infty}^\infty\lim_{N\to\infty}\sqrt{\frac{m\omega}{\pi i\hbar}}
      \frac{e^{-i\frac{M-1}2\pi}}{\sqrt{|\sigma(N)|^2}}
      \exp\biggl[\frac{u^2(x_I)+v^2(x_I)}2\biggr]
      \frac{\exp\Bigl[-\frac{u^2(x_I)}{1+z(N)}\Bigr]}{\sqrt{|1+z(N)|}}
      \frac{\exp\Bigl[-\frac{v^2(x_I)}{1-z(N)}\Bigr]}{\sqrt{|1-z(N)|}}
      f(x_I)\,dx_I\notag\\
      &=\int_{-\infty}^\infty\lim_{N\to\infty}\sqrt{\frac2{i\pi}}
      \frac{e^{-i\frac{M-1}2\pi}}{\sqrt{|\sigma(N)|^2}}
      \exp\biggl[\frac{u^2(x_I(v))+v^2}2\biggr]
      \frac{\exp\Bigl[-\frac{u^2(x_I(v))}{1+z(N)}\Bigr]}{\sqrt{|1+z(N)|}}
      \frac{\exp\Bigl[-\frac{v^2}{1-z(N)}\Bigr]}{\sqrt{|1-z(N)|}}
      f(x_I(v))\,dv
      \label{effhon-eq:21}
    \end{align}
    where we have made a change of the integral variable $x_I$ to $v$.
    Making use of the formula
    \begin{equation}
      \label{effhon-eq:22}
      \int_{-\infty}^\infty e^{-wx^2}f(x)\,dx
      =e^{-i\frac12\arg w}\int_{-\infty}^\infty e^{-|w|x^2}f(xe^{-i\frac12\arg w})\,dx
    \end{equation}
    with $w\in\mathbb{C}$, $|\arg w|<\pi/2$, we rewrite (\ref{effhon-eq:21}) to
    \begin{equation}
      \label{effhon-eq:23}
      (\ref{effhon-eq:21})
      =\lim_{N\to\infty}e^{-i\frac12\arg\frac1{1-z(N)}}\int_{-\infty}^\infty
      \sqrt{\frac2{i\pi}}\frac{e^{-i\frac{M-1}2\pi}}{\sqrt{|\sigma(N)|^2}}
      e^{\frac{u^2+v^2}2}\frac{e^{-\frac{u^2}{1+z(N)}}}{\sqrt{|1+z(N)|}}     
      \frac{e^{-\frac{v^2}{|1-z(N)|}}}{\sqrt{|1-z(N)|}}f(x_I)\,dv
    \end{equation}
    where $v$ in arguments has become $ve^{i\frac12\arg(1-z)}$.
    By using one of the expression of the $\delta$-function
    \begin{equation*}
      \lim_{r\to\infty}\frac1{\sqrt{\pi r}}e^{-\frac{x^2}r}=\delta(x),
    \end{equation*}
    with $r=|1-z(N)|$, we obtain
    \begin{equation}
      \label{effhon-eq:24}
      (\ref{effhon-eq:23})=\int_{-\infty}^\infty e^{-i\frac M2\pi}\delta(v)f(x_I(v))\,dv
    \end{equation}
    where $x_I(v)$ is $x_I(ve^{-i\frac\pi4})$ actually, 
    but because of the nature of the $\delta$-function, 
    $x_I(ve^{-i\frac\pi4})$ is the same with $x_I(v)$.
    Making a change of variable $v$ to $x_I$, we finally obtain
    \begin{equation*}
      (\ref{effhon-eq:24})=\int_{-\infty}^\infty e^{-i\frac M2\pi}\delta(x_F-x_I)f(x_I)\,dx_I
    \end{equation*}
    so we find
    \begin{equation}
      \label{effhon-eq:25}
      K(x_F,x_I;T)=e^{-i\frac M2\pi}\delta(x_F-x_I)=e^{-i\frac M2\pi}\delta(x_F-(-1)^Mx_I)   
    \end{equation}
  \item $M\in\mathbb{N}\setminus2\mathbb{N}$
    
    In this case
    \begin{equation*}
      1+z(N)=2\sin\varepsilon(N)e^{-i(\frac\pi2-\varepsilon(N))}
    \end{equation*}
    and
    \begin{equation*}
      z(N)\to-1,\quad\arg(1+z(N))=-\frac\pi2\qquad(N\to\infty).
    \end{equation*}
    In the similar way with (\ref{effhon-item:3}), we obtain
    \begin{equation}
      \label{effhon-eq:26}
      K(x_F,x_I;T)=e^{-i\frac M2\pi}\delta(x_F+x_I)=e^{-i\frac M2\pi}\delta(x_F-(-1)^Mx_I)
    \end{equation}
  \end{enumerate}
  Putting (\ref{effhon-eq:25}) and (\ref{effhon-eq:26}) together, we obtain
  \begin{equation}
    \label{effho-eq:9}
    K(x_F,x_I;T)=e^{-i\frac M2\pi}\delta(x_F-(-1)^Mx_I)\qquad(M\in\mathbb{N}\cup\{0\})
  \end{equation}
  This is in accordance with (\ref{effho-eq:2}).
\end{enumerate}

\section{Discussion}
\label{effhon-sec:discussion}

In this paper we have derived the extended Feynman formula by the 
discrete time formulation of path integral.
As stated in \cite{Schulman}, we can observe clearly that when time passes over
every caustic point, number of negative eigenvalues increases and the 
phase correction is multiplied in the Feynman formula.

As pointed out in \cite{Horvathy2}, if we integrate (\ref{effhon-eq:7})
formally to
\begin{equation*}
  \frac1{\sqrt{\det A}}=\frac1{\sqrt{(-1)^M|\det A|}},
\end{equation*}
we do not know which branch of $\sqrt{-1}$ should be chosen.
The Frensel integral seems to be indispensable to obtain correct number of 
eigenvalues in each sign.

The extensions of the Feynman formula for other systems like a
forced harmonic oscillator have been made \cite{cheng1,cheng2,cheng3,miller1,miller2,marcus}.
The extensions by the discrete time method will be applicable
to these systems.

\begin{center}
  \large\bfseries Acknowledgments
\end{center}
We thank K. Fujii for important ideas and useful discussions.

\end{document}